\documentclass [aps,pra,twocolumn,amsfonts]{revtex4}
\usepackage{graphicx}
\newcommand {\beq}{\begin{equation}}
\newcommand {\eeq}{\end{equation}}
\newcommand {\beqa}{\begin{eqnarray}}
\newcommand {\eeqa}{\end{eqnarray}}

\begin{document}

\title{Entanglement enhanced classical capacity of quantum communication
channels with memory in arbitrary dimensions}

\author{E.~Karpov, D.~Daems, and N.~J.~Cerf}

\affiliation{Quantum Information and Communication, \'Ecole
Polytechnique, CP 165, Universit\'e Libre de Bruxelles, 1050
Brussels, Belgium}

\begin{abstract}

We study the capacity of $d$-dimensional quantum channels with memory modeled by correlated noise.
We show that, in agreement with previous results on Pauli qubit channels,
there are situations where maximally entangled input states achieve higher values of
mutual information than product states.
Moreover, a strong dependence of this effect on the nature of the noise correlations
as well as on the parity of the space dimension is found. We conjecture that when entanglement
gives an advantage in terms of mutual information,
maximally entangled states saturate the channel capacity.

\end{abstract}

\maketitle\texttt{}

\section{Introduction}


One of the major problems of quantum information is the evaluation
of the capacity of quantum communication channels, i.e., the
evaluation of the amount of \emph{classical information} which can
be reliably transmitted by \emph{quantum states}. This problem has been
extensively studied along with the development of other aspects of
quantum information science. Early works in this direction were devoted,
mainly, to memoryless channels for which consecutive signal
transmissions through the channel are not correlated. The
capacities of some of such channels were determined
\cite{H73}-\cite{BF05}
and it was proven that in most cases their capacities are
additive. For Gaussian memoryless channels under Gaussian inputs,
the multiplicativity of output purities was
proven in  \cite{SEW05} and the additivity of the
energy-constrained capacity even in the presence of
classical noise and thermal noise was proven in \cite{Hiro05}.

In the last few years, much attention was given to quantum channels
with memory \cite{Mac02}-\cite{G05}
with the hope to find a possibility to enhance the channel capacity
by using entangled input states.
This would be possible if the capacity of such
channels is superadditive. For some of such channels, for example,
for bosonic memory channels in the absence of input energy
constraints, the additivity conjecture was proven leaving no hope
to enhance channel capacity using entangled inputs \cite{GM05}.
However, for several other examples of quantum channels, where
memory is introduced by a correlated noise, it was shown that
entangling two consecutive uses of the channel enhances the
overall channel capacity.
These examples include qubit Pauli channels \cite{Mac02,Mac04}
and bosonic continuous-variable Gaussian channels \cite{CCMR05}-\cite{GM05}.
For qubit channels it was shown that
if the noise correlations are stronger than some critical value,
maximally entangled input states enhance the channel capacity compared
to product input states.
For Gaussian channels also, entangled states perform better than product states.
However, for each value of the noise correlation parameter,
there exists an optimal degree of entanglement (and not maximal entanglement)
that maximizes the channel capacity.
Quantum channels with correlated noise in dimensions $d>2$
were not considered in the literature in this
context yet (see Note added).
They correspond to a kind of intermediate systems between the qubit
and Gaussian channels.
Therefore, we expect to find new features that this
intermediate dimensionality can add to the known facts.
We start with a short review of the results on the capacities of the Pauli
qubit channels with memory studied in papers \cite{Mac02,Mac04}.
Then, we propose a generalization of these channels to $d$ dimensions and
present new results on their capacity.

\section{Capacity of qubit channels with correlated noise}

The action of a transmission channel on an initial state $\rho$
is given by a completely positive (CP) map ${\mathcal E}$
\begin{equation}
  \rho \rightarrow {\mathcal E}(\rho).
\end{equation}
The amount of classical information which can be reliably transmitted
through a quantum channel is given by the Holevo-Schumacher-Westmoreland
bound \cite{H73,SW97} as the maximum of mutual information
\begin{equation}\label{HSW}
   \chi(\mathcal E)
        = \max_{\{P_i,\rho_i\}} I(\mathcal E)
\end{equation}
taken over all possible ensembles $\left\{P_i,\rho_i\right\}$
of input states $\rho_i$ with {\it a priori} probabilities
$P_i\ge 0, \quad \sum_i P_i =1$.
The mutual information of an ensemble $\left\{P_i,\rho_i\right\}$ is defined as
\begin{equation}\label{I}
   I(\mathcal E(\{P_i,\rho_i\}))
     =  \left[ S\left(\sum_i P_i\mathcal E(\rho_i)\right)
                     -\sum_i P_i S(\mathcal E(\rho_i))
                \right]
\end{equation} where $S(\rho) = - {\rm Tr} [\rho \log_2\rho]$ is
the von Neumann entropy.

If we find a state $\rho_*$ which minimizes the output entropy
$S({\mathcal E(\rho_*)})$ and replace the first term in (\ref{I})
by the largest possible entropy given by the entropy of the
maximally mixed state, we obtain the following bound
\begin{equation}\label{bound}
  \chi({\mathcal E})\le \log_2(d)-S({\mathcal E}(\rho_*))
\end{equation}
where  $\rho$ is a $d\times d$ matrix.
This bound (\ref{bound}) is very useful for further evaluation
because it was shown to become tight for two-qubit channels
\cite{Mac04} which are covariant under Pauli rotations, $\sigma_i$,
such that the following equality holds:
\begin{equation}\label{cov}
 {\mathcal E}_2(\sigma_i\otimes\sigma_j\rho\sigma_i\otimes\sigma_j)
     =\sigma_i\otimes\sigma_j{\mathcal  E}_2(\rho)\sigma_i\otimes\sigma_j.
\end{equation}
The two-qubit channels represent two consecutive applications  (denoted by ${\mathcal E}_2$)
of a one-qubit channel.

The proof of the above statement is simple. It is based on the fact that the Pauli
matrices form an irreducible representation of a Heisenberg group.
This implies that for any two-qubit state $\rho$, an equal probability ensemble
\begin{equation}\label{ens}
  \frac{1}{16}\sum_{i,j=0}^{3}\sigma_i\otimes\sigma_j\rho\sigma_i\otimes\sigma_j
\end{equation}
is maximally mixed. Now, let $\rho_*$ be an input
two-qubit state that minimizes the output entropy $S({\mathcal E}_2(\rho))$.
Then, the equal probability ensemble (\ref{ens}) made
from this $\rho_*$ has the following properties. On the one hand,
it is maximally mixed and therefore, it maximizes the first term
in (\ref{I}). On the other hand, due to the
channel covariance (\ref{cov}), the second term in (\ref{I}) becomes
equal to the entropy $S({\mathcal E}(\rho_*))$ which is minimal by our choice of
$\rho_*$. Therefore, the equal probability ensemble (\ref{ens})
made from $\rho_*$ attains the bound (\ref{bound}). Finally,  in
order to find $\chi({\mathcal E}_2)$ for covariant two-qubit
channels (\ref{cov}) it is enough to find one of the
\emph{optimal} states $\rho_*$ and calculate the right hand side
of the bound (\ref{bound}), which is tight in this case. We will
present a $d$-dimensional version of these arguments in the next
section.

In general, being a CP map, any quantum channel 
can be represented by an operator-sum:
\begin{equation}
 {\mathcal E}(\rho)= \sum_k A_k\rho A_k^\dag,
                     \qquad \sum_k A_k^\dag A_k= \openone .
\end{equation}
When ${\mathcal E}_n$ represents $n$ uses of the same quantum channel,
the Hilbert space of the initial states is a tensor product such that
$\rho \in \mathcal H^ {\otimes n}$.
The amount of information transmitted per use of the channel in the 
limit of infinitely many uses of the channel determines the
channel capacity
\begin{equation}
  C=\lim_{n\rightarrow\infty}\frac{1}{n}\sup_{\{P_i,\rho_i\}}I({\mathcal E}_n).
\end{equation}

If consecutive uses of the channel are not correlated then
the repeated uses ($n$ times) of this memoryless channel can be represented
by a tensor product of the form
\begin{equation}\label{product}
  {\mathcal E}_n(\rho)
    = \sum_{k_1\dots k_n} (A_{k_1}\otimes\dots \otimes A_{k_n})\rho
                          (A_{k_1}^\dag\otimes\dots \otimes A_{k_n}^\dag) .
\end{equation}
However, in the presence of correlations between consecutive uses
of the channel the representation given by Eq. (\ref{product}) is
not valid.
As an example, for \emph{Pauli} channels, whose action is represented
by Pauli matrices, $n$ uses of the channel is in general given by
\begin{equation}
    A_{k_1\dots k_n}=\sqrt{p_{k_1\dots k_n}} \sigma_{k_1}\otimes\dots\otimes\sigma_{k_n},
\end{equation}
where the probability distribution $p_{k_1\dots k_n}$ is
normalized $\sum_{k_1\dots k_n}p_{k_1\dots k_n}=1$ and its
particular properties determine different kinds of $n$-qubit Pauli channels.
A Pauli channel is memoryless and can be represented as in Eq.
(\ref{product})  when $p_{k_1\dots k_n}$ is factorized into
probabilities which are independent for each use of the channel,
\begin{equation}\label{markov}
   p_{k_1\dots k_n}=p_{k_1}\dots p_{k_n}.
\end{equation} 
However, if each use of the channel depends on the
preceding one, such that $p_{k_1\dots k_n}$ is given by a product
of conditional probabilities
\begin{equation}\label{pn}
  p_{k_1\dots k_n}=p_{k_1}p_{k_2|k_1}\dots p_{k_n|k_{n-1}}
\end{equation} 
a memory effect is introduced. 
Indeed, the correlations between the consecutive uses of the
channel act as if the channel ``remembers'' the previous signal and
acts on the next one using this ``knowledge''. This type of
channels is called a Markov channel as the probability (\ref{pn})
corresponds to a Markov chain of order 2.

In the two-qubit Pauli channels considered in \cite{Mac02, Mac04}
\begin{equation}\label{mtqc}
    {\mathcal E}_2(\rho)
    = \sum_{i=0}^3p_{ij}\,\sigma_i\otimes\sigma_j\rho\,\sigma_i\otimes\sigma_j , \qquad
      \sum_{i,j=0}^3p_{ij}=1
\end{equation}
a memory effect is introduced by choosing
probabilities $ p_{ij}$ which include a correlated noise
represented by the term with a Kronecker's delta:
\begin{equation}\label{p}
  p_{ij}=q_iq_{i|j}=q_i((1-\mu)\,\,q_j+\mu\, \delta_{ij}).
\end{equation}
The memory parameter $\mu\in[0,1]$ characterizes the correlation
``strength''. Indeed, for $\mu = 0$, the probabilities of two
subsequent uses of the channel are independent, whereas for $\mu=1$, the
correlations are the strongest ones.
These channels are not exactly Markov for the following reason.
Let us send $2n$ qubits through such channels one after another
and consider $n$ consecutive pairs of these qubits.
The actions of the channel on the qubits belonging to the same pair
are correlated according to (\ref{mtqc}) and (\ref{p}).
However, the actions of the channel on
the qubits from different pairs are uncorrelated so that
$p_{k_1\dots k_{2n}}=p_{k_1}p_{k_2|k_1}\dots p_{k_{2n-1}}p_{k_{2n}|k_{2n-1}}$.
Therefore $2n$ consecutive uses of the channel are factorized
into pairs and  can be represented by a product of two-qubit channels 
${\mathcal E}_{2n}= {\mathcal E}_2^{\otimes n}$
where ${\mathcal E}_2$ is defined by (\ref{mtqc}).
However, even these ``limited'' pairwise correlations
lead to the advantages of using entangled input states 
as it was shown in \cite{Mac02,Mac04}.

\vskip 0.5cm
\emph{A. Quantum depolarizing (QD) channel}.
This channel is determined by equal probabilities for all Pauli matrices
except for the identity:
      \begin{equation}\label{p1}
            q_0  =  p, \quad
            q_1 = q_2 = q_3   =  q=(1-p)/3.
      \end{equation}
It is characterized by the parameter $\eta=p-q=(4p-1)/3 \in [-1/3,1]$, which is called 
``shrinking factor'' \cite{Mac02}.

\vskip 0.5cm
\emph{B. Quasi-classical depolarizing (QCD) channel}.
This channel is determined by a probability parameter
which has the same value $p$ for the two Pauli matrices
which do not introduce any bit-flip and another value $q$
for the other two Pauli matrices including a bit flip:
  \begin{equation}\label{p2}
                q_0 = q_1 = p, \quad
                q_2 = q_3 = q=(1-2p)/2.
  \end{equation}
We call this channel quasi-classical because,
it is equivalent to the concatenation of a 
fully-dephasing channel (where the quantum phase is lost)
followed by a classical channel akin to the depolarizing channel,
where the bit is left unchanged with some probability
or is shifted with the complimentary probability.
For this channel there is no ``shrinking factor''
but it is nevertheless characterized
by a single parameter $\eta=2(p-q)=4p-1 \in [-1,1]$. 

The results of the evaluation \cite{Mac02,Mac04} show that product states
of the form $|00\rangle\langle 00|$ maximize the mutual information
of memoryless channels ($\mu=0$)
of both types, whereas maximally entangled states
of the form $|\Phi^+\rangle\langle\Phi^+|$
maximise the mutual information for the
strongest noise correlations ($\mu=1$).
Moreover for each channel there exists a crossover point $\mu_c$
below which ($0\le \mu<\mu_c$) the mutual information is maximized
by product states  and above which ($\mu_c<\mu\le 1$)
maximally entangled states maximise the mutual information.
This crossover point $\mu_c$ depends on the probability parameters $p$
in (\ref{p1}) and (\ref{p2}).

In Fig. 1, we plot the  mutual information of equal probability ensembles of the type
(\ref{ens}) for product states and for maximally entangled states
in the QCD channel ($\eta=0.4$).
Following the above arguments we evaluate the mutual information
according to the right hand side of Eq. (\ref{bound})
using a product state or a maximally entangled state for $\rho_*$
This plot is qualitatively the same as the one
presented in \cite{Mac02} for the quantum depolarizing channel.

\begin{figure}[ht]
\begin{center}
  \includegraphics[width=0.45\textwidth]{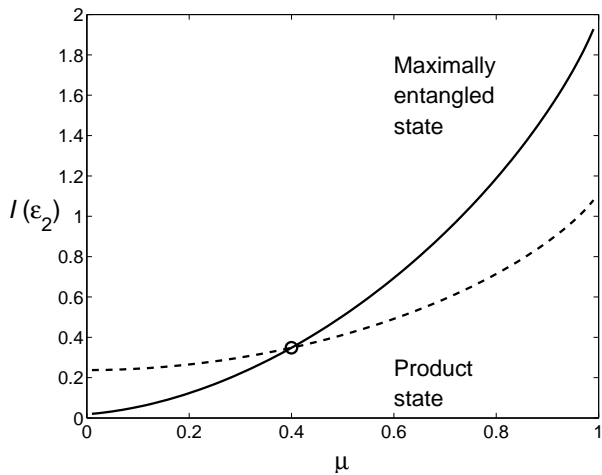}
  \caption{Mutual information $I({\mathcal E}_2)$ as a function of the memory
          parameter $\mu$ for QCD channel ($d=2$) with $\eta=0.4$.
          The solid line corresponds to maximally entangled input states while
          the dashed line corresponds to product states.}
\end{center}
\end{figure}

\section{Quantum channel with correlated noise in d dimensions}

Pauli qubit channels are generalized to $d$-dimensional Heisenberg channels \cite{Cerf00}
constructed with the help of ``error'' or ``displacement'' operators acting
on $d$-dimensional states.
  \begin{equation}\label{U}
    U_{m,n}
       = \sum_{k=0}^{d-1}e^{\frac{2\pi i}{d} kn}\,|k+m\rangle\langle k|.
  \end{equation}
Here the index $m$ characterises a cyclic shift in the computational basis
by analogy with the bit-flip, and the index $n$ characterises a phase shift.
The displacement operators form a Heisenberg group \cite{F95} with commutation relation
\begin{equation}\label{cr}
 U_{m,n}U_{m',n'}
    = e^{2\pi i (m'n-mn')/d}U_{m',n'}U_{m,n}
\end{equation}

Two uses of a $d$-dimensional channel are described by
  \begin{eqnarray}\label{emn}
    {\mathcal E}_2(\rho)
     & = & \sum_{m,n,m',n'=0}^{d-1}p_{m,n,m',n'}\\ \nonumber
     & \times &    (U_{m,n}\otimes U_{m',n'})\,\rho\,
         (U_{m,n}^\dag\otimes U_{m',n'}^\dag).
  \end{eqnarray}
By using the notation ${\mathcal E}_2(\rho)$ we emphasize again
that (\ref{emn}) represents two consecutive uses of a channel
each acting on $d$-dimensional states. For simplicity,
we shall call these channels $d$-dimensional.

We introduce a correlated noise by a Markov-type probability
 \begin{eqnarray}\label{pmn}
    p_{m,n,m',n'}
      & = & (1-\mu)q_{m,n}\,q_{m',n'}\\
      & + & \mu\,q_{m,n}\,\delta_{m,m'}\,
       ((1-\nu)\,\delta_{n,n'}+\nu\,\delta_{n,-n'})\nonumber
  \end{eqnarray}
Here $\mu$ is the memory parameter like in the 2-dimensional case,
but the Kronecker's delta $\delta_{ij}$ in (\ref{p})
is represented now by a product of two Kronecker' deltas
representing the noise correlations
separately for displacements (index $m$) and for phase shifts (index $n$).
In addition, we introduce both phase correlations ($\delta_{n,n'}$)
as well as phase anticorrelations ($\delta_{n,-n'}$)
with a new parameter $\nu$ characterising the type of
phase correlations in the channel.
For $d=2$ such a distinction disappears
as phase correlations $\delta_{n,n'}$
and phase anticorrelations $\delta_{n,-n'}$ coincide,
but, interestingly for higher dimensions the type of phase correlations
affects the channel capacity. In the limiting case of infinite
dimensional bosonic Gaussian channels, only phase anticorrelations
 were shown to provide some enhancement 
of the channel capacity due to entanglement \cite{CCMR05}.

The examples of two-qubit channels presented above
are generalized to $d$-dimensions as follows:

\vskip 0.5 cm
\emph{A. Quantum depolarizing (QD) channel}.
This channel is given by the following probability parameters
which are equal to each other for all possible actions of the channel except for 
the identity as in the 2-dimensional case.
\begin{equation}\label{qdc}
       q_{m,n} = \left\{\begin{array}{ll}
                                      p & , \quad m=n=0 , \\
                       {\displaystyle q= \frac{1-p}{d^2-1}} & , \quad \textnormal{otherwise}.
                         \end{array}
                 \right.
   \end{equation}
The sum of all probabilities is normalised therefore, $p$ and $q$ are not independent
and the channel may be characterized by a single parameter $\eta=p-q$ with the range 
$\eta \in [-1/(d^2-1),1]$.
This reminds us the``shrinking factor'' for two-qubit QD channel.

\vskip 0.5cm
\emph{B. Quasi-classical depolarizing (QCD) channel}.
This channel is given by the following probability parameters

   \begin{equation}\label{qcdc}
      q_{m,n}= q_m = \left\{\begin{array}{ll}
                                         p  &,\quad m=0, \\
                          {\displaystyle q  =\frac{1-dp}{d(d-1)}} &,\quad \textnormal{otherwise} .
                            \end{array}
                      \right. 
    \end{equation}
The probabilities of the shifts by $m$ are equal
regardless of the phase shift determined by $n$.
Moreover, we have chosen them to be equal if $m>0$.
The probability of ``zero'' displacement ($m=0$)
differs form others as in the 2-dimensional case.
The sum of all probabilities is normalised so that $p$ and $q$ are not independent
and the channel may be characterized by one parameter, 
which we choose to be $\eta=d(p-q)$
because it takes the values in the interval $ [-1/(d-1),1]$, which is close to the 
the``shrinking factor'' for QD channel.

For these  $d$-dimensional channels,  we prove, first, that bound (\ref{bound}) is tight.
Let $\rho_*$ be the input state that minimizes the output entropy $S(\mathcal E_2(\rho_*))$
and let us notice that owing to the commutation relation
(\ref{cr}) the channel determined by (\ref{emn}) and (\ref{pmn})
is covariant with respect to the displacements (\ref{U}):
\begin{eqnarray}\label{cov1}
 \lefteqn{{\mathcal E}_2(U_{m,n}\otimes U_{m',n'}\rho_*U^\dag_{m,n}\otimes U^\dag_{m',n'})}\\
  & = & U_{m,n}\otimes U_{m',n'}{\mathcal E}_2(\rho_*)U^\dag_{m,n}\otimes U^\dag_{m',n'}. \nonumber
\end{eqnarray}
Using the fact that the von Neumann entropy
is invariant under unitary transformations
\begin{equation}\label{rhomn}
 \rho_{mnm'n'}=U_{m,n}\otimes U_{m',n'}\rho_*U^\dag_{m,n}\otimes U^\dag_{m',n'}
\end{equation}
we come to
\begin{equation}\label{SE}
  S({\mathcal E}_2(\rho_{mnm'n'}))=S({\mathcal E}_2(\rho_*)).
\end{equation}
Due to the fact that the group of matrices $\{U_{m,n}\otimes U_{m',n'}\}$
is an irreducible representation of the Heisenberg group,
the ensemble of input states $\rho_{mnm'n'}$
taken with equal probabilities provides a maximally mixed output state,
which up to the factor $1/d^2$ is:
\begin{eqnarray}\label{mm}
  \lefteqn{{\mathcal E}_2\,\left(\sum_{m,n,m',n'=0}^{d}\rho_{mnm'n'}\right)} \nonumber \\
  & = & \sum_{m,n,m',n'=0}^{d^2}U_{m,n}\otimes U_{m',n'}{\mathcal E}_2(\rho_*)U^\dag_{m,n}\otimes U^\dag_{m',n'}\nonumber \\
  & = &\openone .
\end{eqnarray}
Then using equal probability $1/d^2$ we insert (\ref{mm}) into the first term of (\ref{I})
and (\ref{SE}) into the second term of (\ref{I}) and thus conclude
that this ensemble indeed maximises the mutual information (\ref{I})
and provides the capacity
  \begin{equation}\label{Irho}
\chi(\mathcal E_2)
      = I(\mathcal E_2(\rho_*))
      \equiv \log_2(d^2)-S(\mathcal E_2(\rho_*)).
  \end{equation}
Thus we have proven that in order to determine the capacity
of these channels we have to find an \emph{optimal} state $\rho_*$
that minimizes the output entropy $S(\mathcal E_2(\rho_*))$.

By analogy with the two-dimensional case \cite{Mac02,Mac04} we start looking 
for the optimal $\rho_*$ by using as an ansatz for the input state
a pure state $\rho_{\rm in}=|\psi_0\rangle\langle \psi_0|$ where
\begin{equation}\label{iniphi}
  |\psi_0\rangle = \sum_{j=0}^{d-1}\alpha_j e^{i\phi_j}|j\rangle|j\rangle, \quad \alpha_j \ge 0, \quad
   \sum_{j=0}^{d-1}\alpha_j^2 =1.
\end{equation}
This allows us to go from a product state to a maximally entangled state
by changing the parameters $\alpha_j$ and $\phi_j$.
Indeed, the choice $\alpha_j=\delta_{j,0}$ and $\phi_j=0$ results in a product state
whereas the choice $\alpha_j=1/\sqrt{d}$ and $\phi_j=0$ results in a maximally entangled state.
Taking into account the form (\ref{U}) of the displacement operators $U_{m,n}$,
the probability distribution $p_{m,n,m',n'}$ (\ref{pmn})
and the probability parameters $q_{m,n}$ for both channels (\ref{qdc}) and (\ref{qcdc}),
we evaluate the action of the channel given by Eq. (\ref{emn})
on the initial state $|\psi_0\rangle\langle\psi_0|$ in the form (\ref{iniphi}).
Then, we diagonalize the output states and find their von Neumann entropy,
which allows us to obtain the mutual information according to Eq. (\ref{Irho}).

First, we evaluate the action of the QD channel given by Eqs. (\ref{emn}-\ref{qdc})
on a pure initial state given by Eq. (\ref{iniphi}).
The result is given by the following equation
\begin{equation}\label{eqdc}
  {\mathcal E}_2(|\psi_0\rangle\langle\psi_0|)
    = (1-\mu)A + \mu[(1-\nu)B+\nu C+D]
\end{equation}
where factors $A$, $B$, $C$, and $D$ for \emph{even} dimensions $d$ are given by
\begin{eqnarray}\label{qdcev}
  A & = & d^2q^2 \openone + (p-q)^2|\psi_0\rangle\langle\psi_0 |\nonumber \\ [2ex]
    & + & dq(p-q)\sum_{j=0}\alpha_j^2  \nonumber
           \left(I\otimes|j\rangle \langle j| + |j\rangle \langle j|\otimes I\right), \nonumber \\
  B & = & d q \sum_{j,m=0}^{d-1}\alpha_j^2 |j+m \rangle |j+m\rangle \langle j+m| \langle j+m| \nonumber  \\
    & + & d \sum_{j,m=0}^{d-1}\alpha_j\alpha_{j+\frac{d}{2}}
               e^{i\left(\phi_j-\phi_{j+\frac{d}{2}}\right)}\nonumber \\
    & \times &  |j+m \rangle |j+m\rangle
            \left\langle j+m+\frac{d}{2}\right|\left\langle j+m+\frac{d}{2}\right|,\nonumber \\
  C & = & d q \sum_{i,j,m=0}^{d-1} \alpha_i \alpha_j e^{i(\phi_i+\phi_j)} \nonumber \\
    & \times &  |i+m \rangle |i+m\rangle\langle j+m|\langle j+m|,\nonumber \\ [2ex]
  D & = & (p-q)|\psi_0\rangle\langle\psi_0 |.
\end{eqnarray}
For the initial product state ($\alpha_j=\delta_{j,0}$) the output state
given by Eq. (\ref{eqdc}) is diagonal and the eigenvalues are
found easily as

\begin{eqnarray}\label{eigenprod11}
  \lambda_{00} & = & (1-\mu)(dq+p-q)^2+\mu(dq+p-q), \nonumber \\
  \lambda_{m0} & = & \lambda_{0n}=(1-\mu)dq(dq+p-q) , \quad m,n > 0 ,\nonumber \\
  \lambda_{mn} & = & (1-\mu)d^2q^2  + \mu dq \delta_{m,n}, \quad m,n > 0.
\end{eqnarray}
Observing that these eigenvalues do not depend on $\nu$ we conclude that product states
do not feel the difference between the types of phase correlations in the channel.

For the maximally entangled initial state ($\alpha_j=1/\sqrt{d}$ and $\phi_j=0$)
we had to rearrange the $d^2\times d^2$ matrix representing the output state Eq. (\ref{eqdc})
in order to find the eigenvalues

\begin{eqnarray}\label{eigenprod12}
  \lambda_{00} & = & (1-\mu)[p(p-q)+q] \nonumber\\
               & + & \mu(2(1-\nu)dq+\nu d^2q+p-q), \nonumber \\
  \lambda_{m0} & = & (1-\mu)q(1+p-q)+2\mu(1-\nu)dq,\nonumber \\
               &   & 0 < m < \frac{d}{2} \nonumber \\
  \lambda_{m0} & = & (1-\mu)q(1+p-q), \quad \frac{d}{2} \le m < d ,\nonumber \\
  \lambda_{mn} & = & (1-\mu)q(1+p-q), \quad m,n > 0.
\end{eqnarray}
We note that for $d=2$ the dependence on $\nu$ also disappears
and we recover the eigenvalues obtained in \cite{Mac02}.

Observing that the state indexes in $B$ (\ref{qdcev}) contain the term $d/2$,
which cannot exist for $odd$ dimensions,
we expect different results for $odd$ $d$.
Indeed, although the action of the channel on $\rho$
for \emph{odd} dimensions is given by the same
Eqs. (\ref{eqdc}) and (\ref{qdcev}),
 $B$ in this case is different and given by
\begin{equation}\label{qdcodd}
  B  = d q \sum_{j,m=0}^{d-1}\alpha_j^2 |j+m \rangle |j+m\rangle \langle j+m| \langle j+m|.
\end{equation}

Nevertheless, the eigenvalues for the initial product state ($\alpha_j=\delta_{j,0}$) are
given by the same Eq. (\ref{eigenprod11}).
However, for the maximally entangled initial state ($\alpha_j=1/\sqrt{d}$),
rearranging the $d^2\times d^2$ matrix representing the output state given by Eq. (\ref{eqdc})
we find the eigenvalues which differ from the case of $even$ dimensions.
\begin{eqnarray}
  \lambda_{00} & = & (1-\mu)[p(p-q)+q] \nonumber\\
               & + & \mu((1-\nu)(dq+p-q)+\nu) \nonumber \\
  \lambda_{mn} & = & (1-\mu)q(1+p-q)+\mu (1-\nu) d q\delta_{m,n}, \nonumber \\
               &   & m,n > 0.
\end{eqnarray}

The details of the analytical evaluation for quasi-classical depolarizing channel
are similar. They are presented in the appendix.

In section V we will visualise and discuss these analytic results, but before,
in the next section we will give some arguments
whether these results provide the \emph{optimal} $\rho_*$.

\begin{figure}[ht]
\begin{center}
  \includegraphics[width=0.47\textwidth]{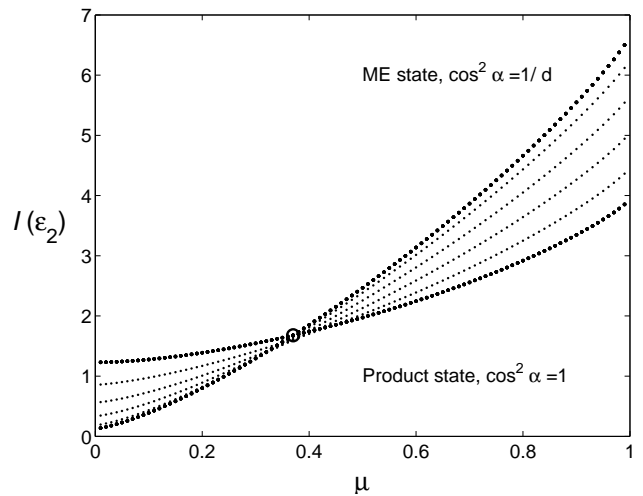}
  \caption{Mutual information $I({\mathcal E}_2(\rho_\alpha))$ as function of the memory  parameter
           $\mu$ for QCD channel with $\eta=0.4$ for different values of the optimisation parameter $\alpha$.
          }
\end{center}
\end{figure}

\section{Discussion on optimization}

The task of finding an \emph{optimal} $\rho_*$
becomes easier for the quasi-classical depolarizing channel
because we can restrict our search from the whole space to a certain subclass.
In order to show this we note, following \cite{Mac04},
that the phase averaging operation ${\mathcal F}$
\begin{equation}
{\mathcal F}(\rho)=\frac{1}{d}\sum_{n=0}^{d-1}
          (U_{0,n}\otimes U_{0,n})\rho(U^\dag_{0,n}\otimes U^\dag_{0,n})
\end{equation}
does not affect the QCD-channel in the sense that
\begin{equation}
   \mathcal E_2\circ{\mathcal F}=\mathcal E_2 .
\end{equation}

Hence, if $\rho_*$ is an \emph{optimal} state
then ${\mathcal F}(\rho_*)$ is also an optimal state.
Therefore we can restrict our search from the whole space ${\mathcal H}^{\otimes 2}$
to ${\mathcal F}({\mathcal H}^{\otimes 2})$.
Finally, using (\ref{U}), it is straightforward to show that
any state  from ${\mathcal F}({\mathcal H}^{\otimes 2})$
is a convex combination of pure states $|\psi_m\rangle\langle\psi_m|$ where
\begin{equation}\label{state}
       |\psi_m\rangle
        =\sum_{j=0}^{d-1} \alpha_j e^{i\phi_i}|j\rangle |j+m\rangle,
          \quad \alpha_j\in {\mathbb R}, \quad \sum^{d-1}_{j=0} \alpha_j^2=1 .
\end{equation}
Restricting our search to the states of the form (\ref{state})
we reduce the number of real optimization parameters from $(2d)^2$ to $2d$,
which can still be a large number.
In order to reduce this number to 1,
we consider the following ansatz
\begin{equation}\label{ansatz}
  |\psi(\alpha)\rangle
         =\cos\alpha|00\rangle + \frac{\sin\alpha}{\sqrt{d-1}}
         \sum^{d-1}_{j=1}|j\, j\rangle ,
  \end{equation}
interpolating between the product state ($\cos\alpha=0$)
and the maximally entangled state ($\cos^2\alpha=1/d$).
Using the one-parameter family of input states $\rho_\alpha=|\psi(\alpha)\rangle\langle\psi(\alpha)|$,
in Fig.~2 we present the mutual information $I(\mathcal E_2(\rho_\alpha))$
for different values of $\alpha$. The mutual information is monotonously modified when $\alpha$ goes
from a product state to a maximally entangled state,
whereas the crossover point $\mu_c$  stays intact.
However, we cannot guarantee that
no other configuration of the parameters $\alpha_i$ and $\phi_i$
minimises the entropy $S(\mathcal E_2(\rho))$ and provides therefore
the maximum of the mutual information.

\section{Main Results}

In this section we will analyse the analytic results
obtained in section III for  product states and maximally entangled states
as candidates for the \emph{optimal} $\rho_*$. 
We first consider the QD channel in \emph{even} dimensions.
In Fig.~3 we display the mutual information $I(\mathcal E_2)$ as a function
of the memory parameter $\mu$ for product states and maximally entangled states.
These curves are seen to cross.
We denote the abscissae of the crossing points as $\mu_c$.
For small noise correlations, $\mu<\mu_c$,
product input states provide higher mutual information 
and for higher noise correlations $\mu>\mu_c$ maximally entangled states do,
as is observed in $d=2$.

\begin{widetext}

\begin{figure}[h]
\begin{center}
\flushleft  (a) \hskip 9cm  (b)

\includegraphics[width=0.47\textwidth]{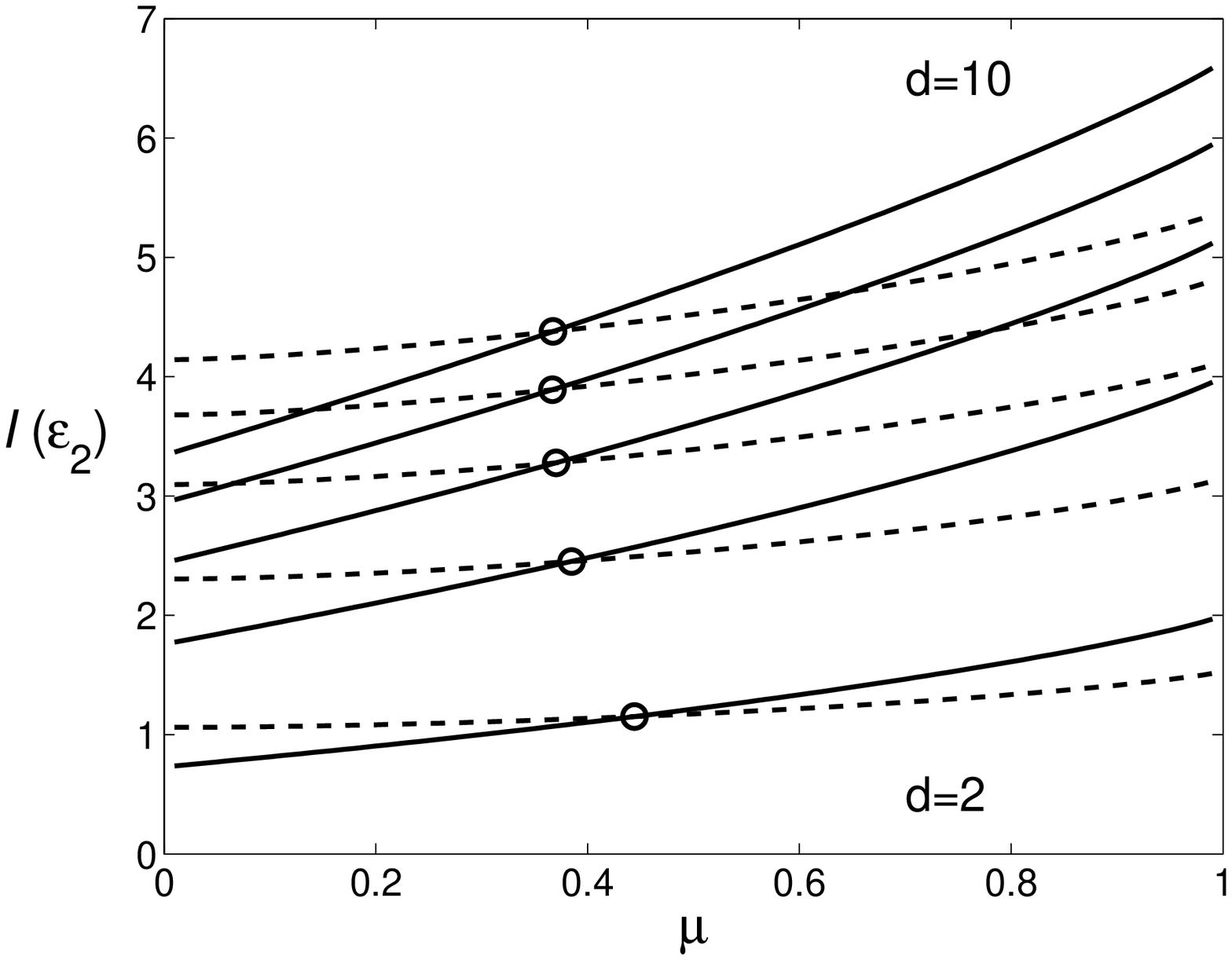}
\hskip 0.7 cm
\includegraphics[width=0.47\textwidth]{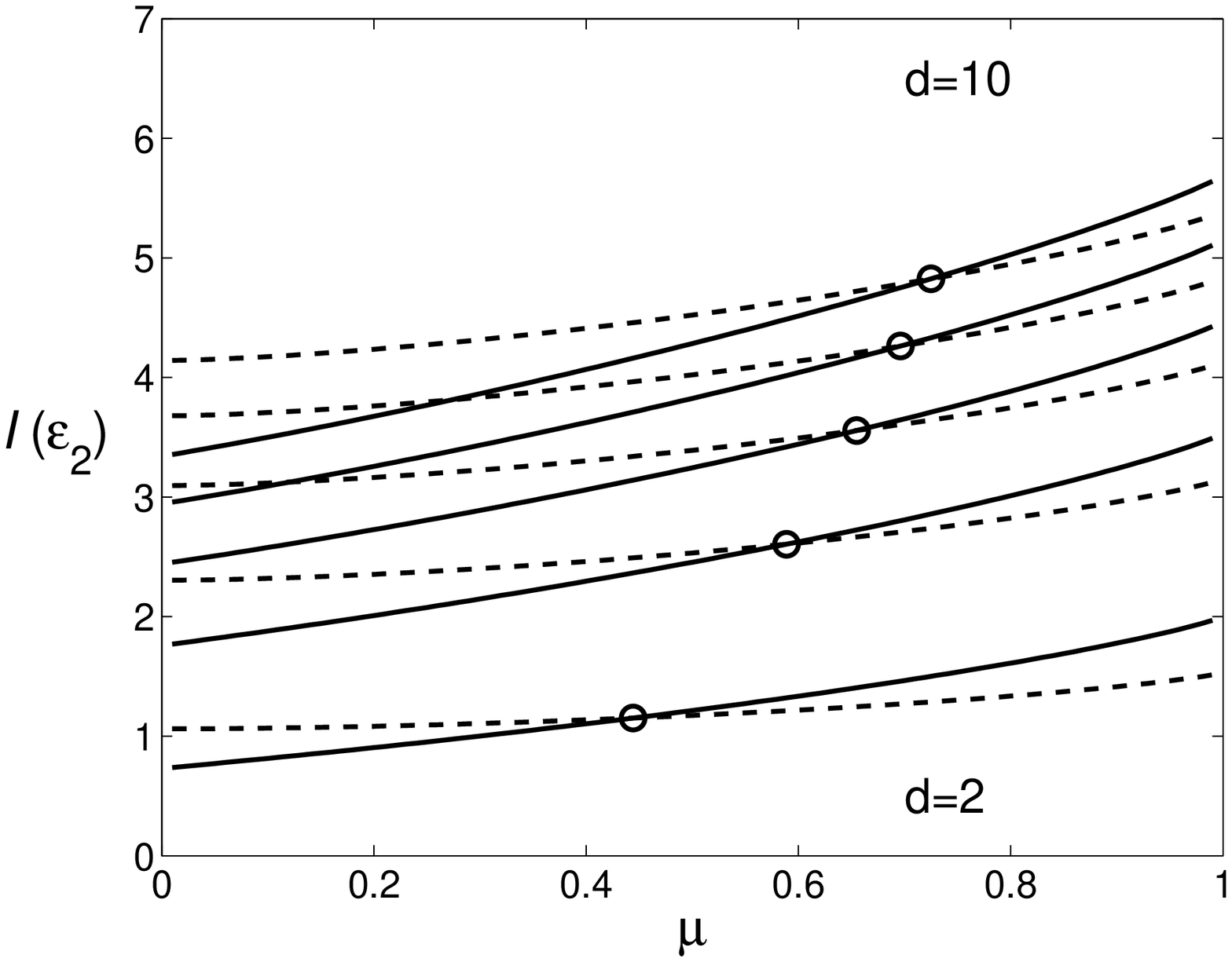}
\caption{Mutual information $I_2({\mathcal E_2}(\rho))$ as function of the memory
         parameter $\mu$ for different dimensions $d=2, 4, 6, 8, 10$
         for the QD entanglement-friendly (a) channel ($\nu=1$)
         and entanglement-non-friendly (b) channel ($\nu=0$)
         both characterized by $\eta=0.8$. The solid lines correspond to maximally entangled
         input states and the dashed lines correspond to the product input states.}
\end{center}
\end{figure}
\end{widetext}

\subsection{Effect of phase correlations}

We now consider how the crossover point $\mu_c$ changes with \emph{even} dimension $d$
for two types of phase correlations in QD channels. Let us notice, first, that
due to the Kronecker's deltas in (\ref{pmn}) the correlated part of the channel
is given by $(U_{m,n}\otimes U_{m,n})\rho (U_{m,n}\otimes U_{m,n})^\dag$ for $\nu=0$
and $(U_{m,n}\otimes U_{m,-n})\rho (U_{m,n}\otimes U_{m,-n})^\dag$ for $\nu=1$.
From the definition of $U_{m,n}$ (\ref{U}) one can see that $n$ determines
the phase shift. The correlations corresponding to the same phase shift $n$
in both entangled signals is what we call ``phase correlations'' (it happens when $\nu=0$)
and the correlations corresponding to opposite phase shifts $n$ and $-n$
is called ``phase anticorrelations'' (it happens when $\nu=1$). For $0<\nu<1$
we have intermediate situations.

In Fig.~3 (a), which corresponds to phase anti-correlations ($\nu=1$),
we see that with an increasing dimension the crossover points move toward smaller $\mu$ thus
widening the interval $]\mu_c,1]$ where maximally entangled states provide
higher values of the mutual information than product states do.
For this reason we call this version of the channel ``entanglement-friendly''.
An opposite effect can be seen in Fig.~3 (b), which corresponds to phase correlations ($\nu=0$).
Here the crossover point moves toward higher values of $\mu$ with an increasing dimension of the space of states,
thus shrinking the interval $]\mu_c,1]$ where maximally entangled states perform better.
This version of the channel is thus called ``entanglement-non-friendly''.
The difference between the two types of phase correlations and therefore,
between the ``entanglement-friendly'' and ``entanglement-non-friendly''
channels may be seen as a result of the fact that maximally entangled states
$(1/\sqrt{d})\sum_k\exp(2\pi i kn/d)|k\rangle|k+m\rangle$ are the eigenstates
of phase anti-correlated products of operators
$U_{m,n}\otimes U_{m,-n}$. Note that for $d=2$
the difference between the two types of phase correlations disappears, which
was discussed in Section II and we recover the result obtained in \cite{Mac02}.

In order to see the effect of the phase correlations for higher dimensions
and observe intermediate situations we draw in Fig.~4 (a)
the coordinate of the crossover point, $\mu_c$, as a function of $d$
for different values of $\nu$.
We see that only strongly anticorrelated phases ($\nu \approx 1$)
provide ``entanglement-friendly'' channels so that with increasing dimension
the interval of $\mu$'s that are favorable for entangled states increases.
In addition, even for $\nu=1$ this increase continues only up to certain $d$,
after which the interval of $\mu$ begins to shrink with increasing $d$.

We also note that $\mu_c$ depends on the shrinking factor $\eta$.
If we drew the curves shown in Fig.~4 (a) for higher values of $\eta$, the slope
of the curves would become steeper and the upper ones, corresponding
to the ``entanglement-non-friendly'' channel ($\nu\approx 0$)
would cross the level $\mu_c=1$ at some $d<100$.
In this case, for higher dimensions, the interval $[\mu_c,1[$ would shrink to zero
so that there would be no values of $\mu$ for which entangled input states may have any advantage at all.

\subsection{QD versus QCD channel}

For the QCD channel the results are similar and therefore we do not present 
the graphs corresponding to Fig.~3 and Fig.~4 (a). Similarly to the picture drawn in Fig.~3,
the anticorrelated phases make the QCD channel also ``entanglement-friendly''
and the correlated phases do the opposite.
The dependence of $\mu_c$ on dimension $d$ corresponding to Fig.~4 (a)
is also qualitatively similar. However, for the QCD channel the upper curves corresponding
to the ``entanglement-non-friendly'' version of the channel ($\nu\approx 0$)
cross the level $\mu_c=1$ at some values of $d<100$ whereas for the QD channel $\mu_c<1$ 
for any $\nu$ for all even dimensions $d<100$ as shown in Fig.~4 (a).

As a result, we conclude that for \emph{even} dimensions,
the advantages of entangled states are more essential for low
(but not always lowest) dimensions, anticorrelated phases,
smaller values of $\eta$ and QD channels.

\begin{widetext}

\begin{figure}[h]
\begin{center}
\flushleft  (a) \hskip 9cm  (b)

\includegraphics[width=0.47\textwidth]{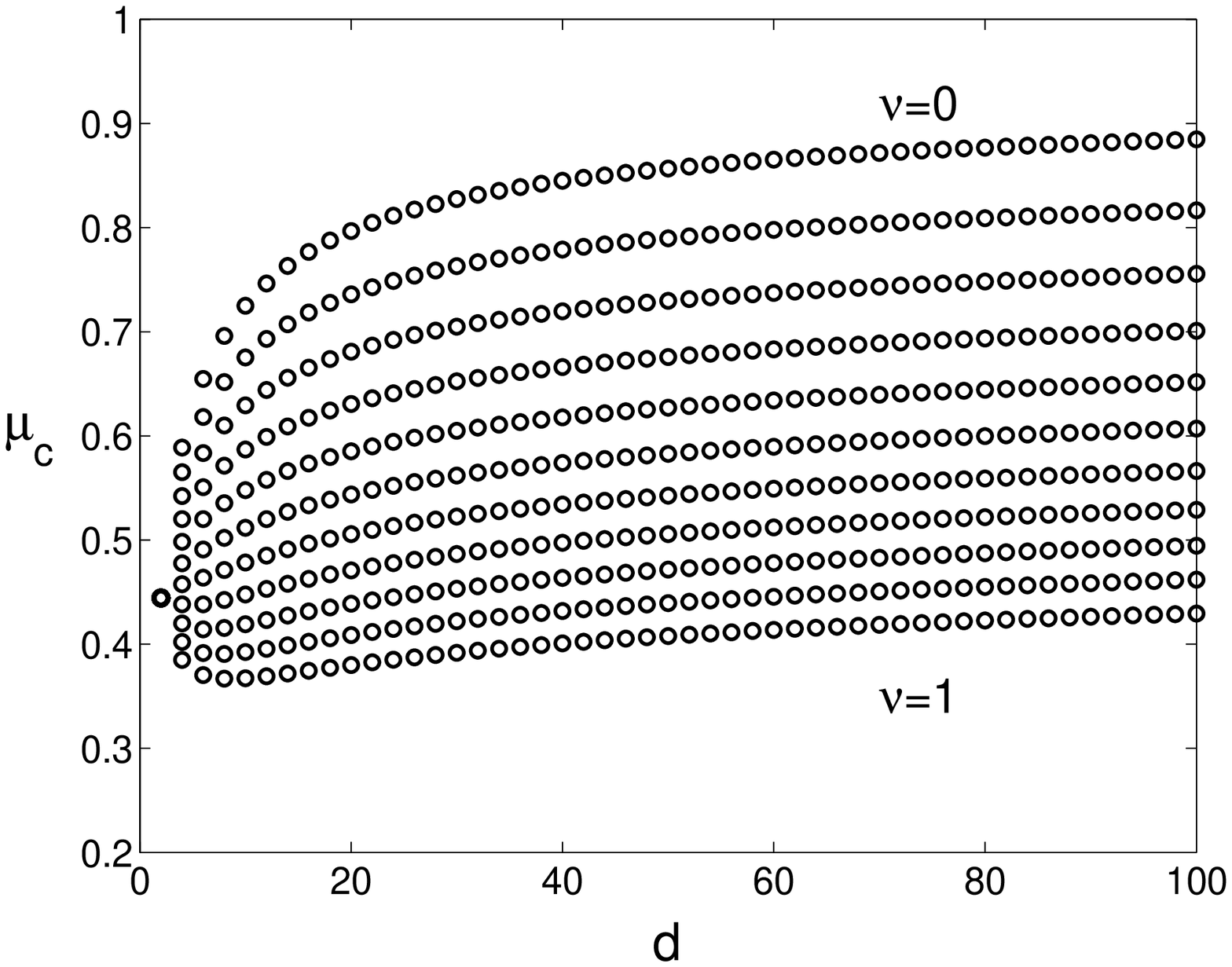}
\hskip 0.7 cm
\includegraphics[width=0.47\textwidth]{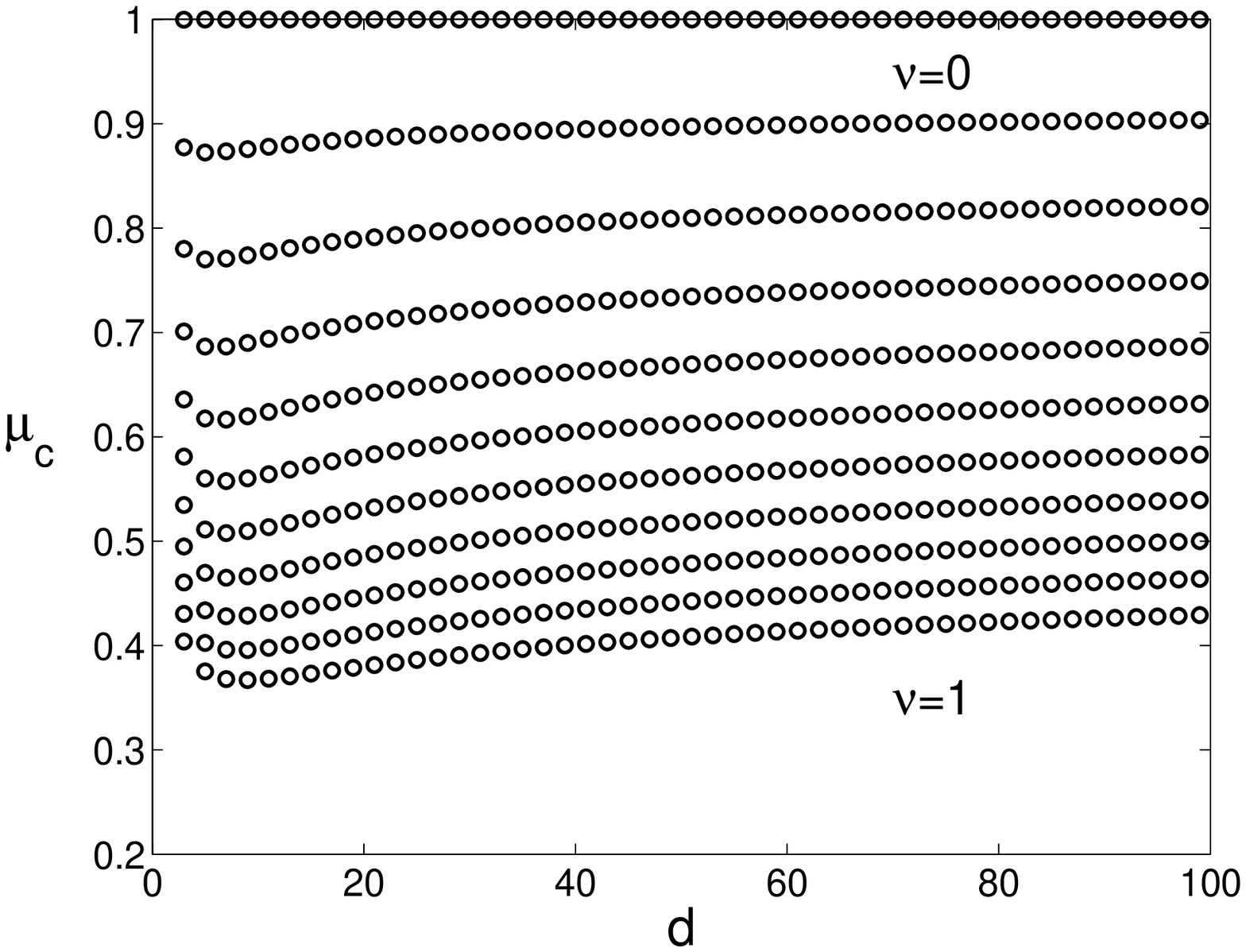}
\caption{Crossover point $\mu_c$ vs. \emph{even} (a) and \emph{odd} (b)
         space dimension for the QD channel with $\eta=0.8$
         for different values of the parameter $\nu$ characterizing phase correlations.
         }
\end{center}
\end{figure}

\end{widetext}

\subsection{Effect of the parity of dimension}

The analytic formulas for \emph{even} dimensions (\ref{qdcev}), (\ref{qcdcev})
and  \emph{odd} dimensions (\ref{qdcodd}), (\ref{qcdcodd}) 
are different. This difference comes from the second term of the expression for $B$ 
in Eqs. (\ref{qdcev}), (\ref{qcdcev}) for \emph{even} dimensions
which is absent in Eq. (\ref{qdcodd}), (\ref{qcdcodd}) for \emph{odd} dimensions.
Due to the factor $(1-\nu)$ in front of $B$ in Eqs. (\ref{eqdc}) and (\ref{eqcdc})
this difference disappears for $\nu=1$, which corresponds to the ``entanglement-friendly'' channels.
Indeed, for the \emph{odd} dimensions, in the case of entanglement-friendly ($\nu=1$) versions
of both QD and QCD channels, the difference from Fig.~3 (a) is only in the position
of the curves whereas qualitatively the curves are similar
and we do not show them here.
However for ``entanglement-non-friendly'' channels ($\nu=1$),
the difference does exist and can be seen from comparison of Fig.~3 (b) (for \emph{even} dimensions)
and Fig.~5 (for \emph{odd} dimensions). Indeed, in Fig.~5 the crossover points
of the entanglement-non-friendly version ($\nu=0$) of the \emph{odd}-dimensional QD channel
lay on the vertical line $\mu=1$. Therefore, effectively there is no crossover
as $\mu$ cannot be larger than 1. In this case for all $\mu$ the maximally entangled
input states do not provide higher values of the mutual information than the product states,
which is not the case for \emph{even} dimensions.

In order to see the effect of intermediate phase correlations, we draw
in Fig. 4 (b) the abscissae  $\mu_c$ of the crossover points, as a function of \emph{odd} $d$
for the QD channel at various values of the ``friendness'' parameter $\nu$.
The upper horizontal line $\mu = 1$ corresponds to the ``entanglement-non-friendly'' 
version of the channel ($\nu=0$) presented also in Fig.~5.
For the QCD \emph{odd}-dimensional channel the picture is similar however,
the upper horizontal line $\mu = 1$ is achieved even for a nonvanishing value
of the ``friendness'' parameter, $\nu =0.3$.

\section{Conclusion}

We have considered two types of $d$-dimensional quantum channels
with a memory effect modeled by a correlated noise.
We have shown that for $d$-dimensional channels with correlated noise
there exist crossover points separating the intervals of the memory parameter $\mu$
where ensembles of maximally entangled input states or product input states provide
higher values of the mutual information.
This result is the same as in the 2-dimensional case.
However it always holds only for channels (which we call ``entanglement-friendly'' channels)
with a particular kind of phase correlations, namely anticorrelations.
For these channels the crossover point move, with increasing $d$, towards
lower values of the memory parameter thus widening the range of correlations
where maximally entangled input states give better results.
For usual phase correlations the situation is opposite, namely,
for higher dimensions of the space the crossover point is shifted towards $\mu_c=1$
so that only for higher degrees of correlations
maximally entangled input states have advantages.
In addition, for these ``entanglement-non-friendly'' channels
the crossover point completely disappears for higher dimensions
so that product input states
always provide higher values of the mutual information
than maximally entangled input states.
Therefore we conclude that the type of phase correlations  strongly affects
this entanglement assisted enhancement of the channel capacity.

\begin{figure}[ht]
\begin{center}
  \includegraphics[width=0.45\textwidth]{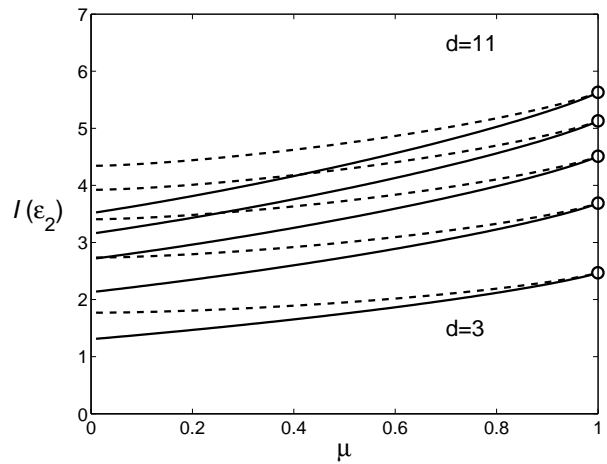}
   \caption{Mutual information $I({\mathcal E}_2(\rho))$ as function of the memory
         parameter $\mu$ for different dimensions $d=3, 5, 7, 9, 11$
         for QD entanglement-non-friendly channel ($\nu=0$) with $\eta=0.8$.}
\end{center}
\end{figure}

In addition, we have observed that the parity of the dimension of the space
of initial states makes an important difference in the
``entanglement-non-friendly'' channels.
Not only the curves of the mutual information vs. the memory parameter
for odd dimensions are shifted with respect to the curves for even dimensions,
but also the channels become completely ``entanglement-non-friendly'' in all odd dimensions
so that maximally entangled input states are always worse than product states.
Strikingly, the channels with anticorrelated noise do not feel the parity of the
space at all. However, any non-vanishing degree of the ``entanglement-non-friendly'' correlations
reveals the parity effect.

We note that the anticorrelated phases remind
the bosonic Gaussian channels considered in \cite{CCMR05}
where the $p$ quadratures are correlated while the $q$ quadratures are anticorrelated.
However, the existence of the crossover point is a significant
difference with the case of the Gaussian channels
for which each value of the noise correlation parameter
determines an optimal degree of entanglement (different from maximal entanglement)
maximising the mutual information.
A challenging problem is to find a link between these results for
$d$-dimensional channels and the results obtained in 
\cite{GM05,CCMR05,RSGM05} for Gaussian channels
with finite energy input signals.

Finally we have presented a parametrisation illustrating a ``monotonous''
deformation of the curves of mutual information vs. the memory parameter,
from product states to maximally entangled states.
The crossover points stay intact during these deformations
which may lead to a threshold type transition maximising
mutual information from product states to maximally entangled states
in case no other states perform better than the maximally entangled ones.
We have to make this stipulation here because 
a full proof of the optimality of maximally entangled input states
is still missing.
{\em Note added:} Recently, we became aware of the work by V.~Karimipour and L. Memarzadeh on channels with memory for $d=3$ \cite{KM06}.

\acknowledgments
We acknowledge the support of the European Union projects
SECOCQ, CHIC and QAP.

\appendix

\section{Quasiclassical depolarising channel}

We evaluate the action  of the channel given by (\ref{emn})
on a pure initial state given by Eq. (\ref{iniphi})
taking into account Eqs. (\ref{U}), (\ref{pmn}), and (\ref{qcdc})
The result is given by the following equation
\begin{equation}\label{eqcdc}
  {\mathcal E}_2(|\phi_0\rangle\langle\phi_0|)
    = (1-\mu) d^2 A + \mu d ((1-\nu)B+\nu C)
\end{equation}
where factors $A$, $B$, and $C$ for \emph{even} dimensions are given by
\begin{eqnarray}\label{qcdcev}
  A & = & q^2 \openone +(p-q)^2 \sum_{j=0}^{d-1} \alpha_j^2|j\rangle |j\rangle \langle j|\langle j| \nonumber\\
    & \times &   q(p-q)\sum_{j=0}^{d-1} \alpha_j^2
                \left(I\otimes|j\rangle \langle j|+|j\rangle \langle j|\otimes I\right), \nonumber \\
  B & = &   \sum_{j=0}^{d-1} \left(q + \alpha_j^2(p-q) \right)
            |j\rangle |j\rangle \langle j| \langle j|\nonumber \\
    & + &  \sum_{j,m=0}^{d-1}\left(q + (p-q)\delta_{m,0}\right)\alpha_j\alpha_{j+d/2}
               e^{i\left(\phi_j-\phi_{j+d/2}\right)} \nonumber \\
    & \times & |j+m \rangle |j+m\rangle
            \left\langle j+m+\frac{d}{2}\right|\left\langle j+m+\frac{d}{2}\right|, \nonumber \\
  C & = &  q \sum_{i,j,m=0}^{d-1} \alpha_i \alpha_j e^{i(\phi_i-\phi_j)}\nonumber \\
    & \times & |i+m \rangle |i+m\rangle\langle j+m|\langle j+m|  \nonumber \\ [2ex]
    & + & (p-q)|\psi_0\rangle\langle\psi_0|.
\end{eqnarray}

For the initial product state ($\alpha_j=\delta_{j,0}$) the output state
given by Eq. (\ref{eqcdc}) is diagonal and the eigenvalues are found easily

\begin{eqnarray}\label{eigenprod21}
  \lambda_{00} & = & (1-\mu)d^2p^2+\mu dp \nonumber \\
  \lambda_{m0} & = & \lambda_{0n}=(1-\mu)d^2pq , \quad m,n > 0 \nonumber \\
  \lambda_{mn} & = & (1-\mu)d^2q^2  + \mu dq \delta_{m,n}, \quad m,n > 0
\end{eqnarray}
Here, as in the case of the quantum depolarizing channel, these eigenvalues
do not depend on $\nu$. Therefore, the product input states
do not feel the difference between
the two types of phase correlations in the channel.

For the maximally entangled initial state ($\alpha_j=1/\sqrt{d}$ )
rearranging the $d^2\times d^2$ matrix
representing the output state given by Eq. (\ref{eqcdc}),
we find the eigenvalues
\begin{eqnarray}\label{eigenprod22}
  \lambda_{00} & = & (1-\mu)d(q^2(d-1)+p^2)+\mu\left((1-\nu)\frac{2}{d}+ \nu\right) \nonumber \\
  \lambda_{m0} & = & (1-\mu)d(q^2(d-1)+p^2) +\mu(1-\nu)\frac{2}{d},\nonumber \\
               &   & 0 < m < \frac{d}{2} \nonumber \\
  \lambda_{m0} & = & (1-\mu)d(q^2(d-1)+p^2) , \quad \frac{d}{2} \le m < d \nonumber \\
  \lambda_{mn} & = & (1-\mu)q(2-d^2q), \quad m \neq n, \quad m,n > 0
\end{eqnarray}
For $d=2$, the dependence on $\nu$ disappears together with the difference
between ``entanglement-friendly'' and ``non-friendly'' channels
and we recover the eigenvalues obtained in \cite{Mac04}.

The result for \emph{odd} dimensions is given by the same Eqs. (\ref{eqcdc}) and (\ref{qcdcev})
with one exception - factor $B$ is given in this case by
\begin{equation}\label{qcdcodd}
  B  =   \sum_{j=0}^{d-1}\left(q+\alpha_j^2(p-q)\right) |j \rangle |j\rangle\langle j|\langle j|. 
\end{equation}

The eigenvalues for the initial product state ($\alpha_j=\delta_{j,0}$) are
given by Eq. (\ref{eigenprod22}).

For the maximally entangled initial state ($\alpha_j=1/\sqrt{d}$)
we had to rearrange the $d^2\times d^2$ matrix representing the output state given by Eq. (\ref{eqcdc})
in order to find the eigenvalues

\begin{eqnarray}
  \lambda_{00} & = & (1-\mu)d(q^2(d-1)+p^2)+\mu\left((1-\nu)\frac{1}{d}+\nu\right) ,\nonumber \\
  \lambda_{m0} & = & (1-\mu)d(q^2(d-1)+p^2) +\mu(1-\nu)\frac{1}{d}, \quad
                      m > 0,  \nonumber \\
  \lambda_{mn} & = & (1-\mu)q(2-d^2q), \quad m \neq n, \quad m,n > 0.
\end{eqnarray}

Again the difference between odd and even dimensions comes form
the second term of the expression for $B$ in Eq. (\ref{qcdcev}) for even dimensions,
which is absent in Eq. (\ref{qcdcodd}) for odd dimensions.
And again, due to the factor $(1-\nu)$ in front of $B$ in Eq.(\ref{eqcdc}),
it is the``entanglement-non-friendly'' phase correlations
that are responsible for the observed parity effect.


\begin{thebibliography}{99}

\bibitem{H73} A.~S.~Holevo, Probl. Inform. Transmission {\bf 9}, 177 (1973)
  (translated from Problemy Peredachi Informatsii);
   IEEE Trans.~Inf.~Theory {\bf 44}, 269 (1998).

\bibitem{SW97} B.~Schumacher and M.~D.~Westmoreland,
   Phys.~Rev.~A {\bf 56}, 131 (1997);  in {\em Quantum Computation and Quantum Information: A Millenium Volume}, edited by S. Lomonaco, Contemporary Mathematics Series (American Mathematical Society, Providence, Rhode Island, 2001).

\bibitem{HSH99} A.~S.~Holevo, M.~Sohma and O.~Hirota,
   Phys.~Rev.~A {\bf 59}, 1820-1628 (1999).

\bibitem{AHW00} G.~G.Amosov, A.~S.~Holevo and R.~F.~Werner,
   Phys.~Rev.~A {\bf 63}, 032312 (2001).

\bibitem{Shor03} P.~W.~Shor,
   quant-ph/0304102, 2003.

\bibitem{GLMSY04} V.~Giovannetti, S.~Lloyd, L.~Maccone, J.~H.~Shapiro and B.~J.~Yen,
   Phys.~Rev.~A {\bf 70}, 032315 (2004).

\bibitem{MPZ01} C. Macchiavello, G. M. Palma and A. Zeilinger,
{\em Quantum Computation and Quantum Information Theory} (World Scientific, Singapore, 2001).

\bibitem{BFMP00} D. Bru\ss , L. Faoro, C. Macchiavello and G. M. Palma, J. Mod. Opt. {\bf 47}, 325 (2000).

\bibitem{KR01} C. King and M. B. Ruskai, IEEE Trans. Inf. Theory {\bf 47}, 192 (2001).

\bibitem{EW05} J.~Eisert and M.~M.~Wolf,
   in {\em Handbook of Nature-Inspired and Innovative Computing} (Springer, New York, 2006).

\bibitem{A05} G.~G.~Amosov,
   Probl. Inf. Transm. {\bf 42}, 67 (2006).

\bibitem{DH05} N.~Datta and A.~S.~Holevo,
   Quantum Inf. Process. {\bf 5}, 179 (2006).

\bibitem{BF05} A.~Sen(De), U.~Sen, B.~Gromek, D.~Bru\ss and M.~Lewenstein,
  Phys. Rev. Lett. {\bf 95}, 260503 (2005).
  
\bibitem{SEW05} A.~Serafini, J.~Eisert and M.~M.~Wolf,
   Phys.~Rev.~A {\bf 71}, 012320 (2005).

\bibitem{Hiro05} T.~Hiroshima,
   Phys. Rev. A {\bf 73}, 012330 (2005).

\bibitem{Mac02} Ch. Macchiavello and G.~M.~Palma,
   Phys. Rev. A {\bf 65}, 050301(R) (2002).

\bibitem{Mac04} Ch.~Macchiavello, G.~M.~Palma and S.~Virmani,
   Phys.~Rev.~A {\bf 69}, 010303(R) (2004).

\bibitem{DWM03} S.~Daffer, K.~W\' odkievicz and J.~K.~McIver,
   Phys.~Rev.~A {\bf 67}, 062312 (2003).

\bibitem{DWCM04} S.~Daffer, K.~W\' odkievicz, J.~D.~Cresser and J.~K.~McIver,
   Phys.~Rev.~A {\bf 70}, 010304 (2004).

\bibitem{BM04} G.~Bowen and S.~Mancini,
   Phys.~Rev.~A {\bf 69}, 012306 (2004).

\bibitem{BDM05} G.~Bowen, I. Devetak and S.~Mancini,
   Phys.~Rev.~A {\bf 71}, 034310 (2005).

\bibitem{CCMR05} N.~J.~Cerf, J.~Clavareau, Ch.~Macchiavello and J.~Roland,
   Phys.~Rev.~A {\bf 72}, 042330 (2005). 

\bibitem{RSGM05} G.~Ruggeri, G.~Soliani, V.~Giovannetti and S.~Mancini,
   quant-ph/0502093.

\bibitem{GM05} V.~Giovannetti and S.~Mancini,
   Phys.~Rev.~A {\bf 71}, 062304 (2005).

\bibitem{G05} V.~Giovannetti,
   J. Phys. A {\bf 38}, 10989 (2005).

\bibitem{KW05} D. Kretschmann and R. F. Werner,
   Phys. Rev A {\bf 72}, 062323 (2005).

\bibitem{KM06} 
 V.~Karimipour and L.Memarzadeh,
    quant-ph/0603223.

\bibitem{AMM05} G.~G.~Amosov, S.~Mancini and V.~I.~Manko,
   J. Phys. A {\bf 39}, 3375 (2006).


\bibitem{Cerf00} N.~J.~Cerf,
   J.~Mod.~Opt. {\bf 47}, 187 (2000).

\bibitem{F95} D.~I.~Fivel,
   Phys.~Rev~Lett. {\bf 74}, 835 (1995).





\end{thebibliography}
\end {document}